\documentclass[epj,final]{svjour}
\usepackage{slashed} 
\usepackage{caption} 
\usepackage{nccmath} 
\usepackage{times} 
\usepackage{amsmath} 
\usepackage{amssymb} 
\usepackage{verbatim} 
\usepackage{graphicx} 
\usepackage{color} 
\usepackage{booktabs} 
\usepackage{subfigure} 
\newcommand{\nths}{\negthickspace\negthickspace} 
\newcommand{\nthn}{\negthinspace\negthinspace}
\newcommand{\lp}{\left(} \newcommand{\rp}{\right)} 
\newcommand{\lb}{\left\{} \newcommand{\rb}{\right\}} 
\newcommand{\ls}{\left[}  \newcommand{\rs}{\right]}
\newcommand{\lv}{\left|}  \newcommand{\rv}{\right|}
\newcommand{\ld}{\left.}  \newcommand{\rd}{\right.}
\DeclareMathOperator{\Ai}{Ai}
\DeclareMathOperator{\J}{J}
\DeclareMathOperator{\Tr}{Tr}

\begin{document} 
\title{High Intensity Compton Scattering in a strong plane wave field of general form}
\author{A Hartin\inst{1} \and G Moortgat-Pick\inst{1,2}}
\institute{DESY, Notkestrasse 85, 22607 Hamburg, Germany \and University of Hamburg, Luruper Chausee 149, 22761 Hamburg, Germany}
\mail{anthony.hartin@desy.de}
\date{}
% PACS: 12.20.Ds
%\ead{anthony.hartin@desy.de}
%\author{A. Hartin, DESY FLC, Notkestrasse 85, Hamburg, Germany}
\abstract{ 
Photon emission by an electron embedded in a strong external field of general form is studied theoretically. The external field considered is a plane wave electromagnetic field of any number of components, period and polarisation. Exact, Volkov solutions of the Dirac equation with the 4-potential of the general external field are obtained. The photon emission is considered in the usual perturbation theory using the Volkov solutions to represent the electron. An expression for the transition probability of this process is obtained after the usual spin and polarisation sums, trace calculation and phase space integration. The final transition probability in the general case contains a single sum over contributions from external field photons, an integration over one of the phase space components and the Fourier transforms of the Volkov phases. The validity of the general expression is established by considering specific external fields. Known specific analytic forms of the transition probability are obtained after substitution of the 4-potential for a circularly polarised and constant crossed external field. As an example usage of the general result for the transition probability, the case of two circularly polarised external fields separated by a phase difference is studied both analytically and numerically.
 % end of PACS codes
} 
\maketitle
 
\section{Introduction} 

The High Intensity Compton Scattering (HICS) is the radiation of a single photon by an electron in an external electromagnetic field. If the external field is sufficiently intense, multiphoton interactions between the electron and the external field become likely and many terms in the usual perturbation series describing the overall interaction, contribute to the final transition probability. Consequently, an interaction picture in which the interaction with the external field, considered to be classical, is calculated exactly and in which the interaction with the quantised boson field is perturbative - the Furry picture - is usually employed \cite{Furry51}.\\

The HICS process is currently the subject of much study largely because new experimental facilities are being planned which will be able to test specific non-linear features of the theory \cite{ELIgrand2009,XFELTDR07}. A variety of experimental conditions can be created each of which require a specific analytic form of the HICS transition probability in order to study. The aim of this paper is to provide a shortcut to this theoretical work by generalising the theoretical description of the transition probability of the HICS process and to illustrate its specific usage in cases of practical interest. \\

Much theoretical work on the HICS process, using the Furry picture and Volkov solutions, was spurred by the construction of the first LASER in 1960 and the possibility of laboratory tests of the theory. The transition amplitude of the HICS process was found to be dependent on the state of polarisation of the external field. For the case of a linearly polarised external electromagnetic field, the HICS transition probability contained an infinite summation of complicated functions which were evaluated in limiting cases only \cite{NikRit64a}. In contrast, a circularly polarised external field results in a transition probability containing Bessel functions, the properties of which are well known \cite{NarNikRit65,BroKib64}. A circularly polarised external field introduces an azimuthal symmetry into the HICS process which results in an analytically less complicated HICS transition probability \cite{Mitter75}. The HICS process for the case of an elliptically polarised external electromagnetic field and for the case of two orthogonal, linearly polarised fields was considered by \cite{Lyulka75}. \\

An important effect emerging from all this theoretical work was a dependency of the energy of the radiated photon (a frequency shift) on the intensity of the external field \cite{BroKib64,Goldman64}. The existence of this frequency shift mechanism allows the HICS process to be used as a generator of high energy photons \cite{Milburn63,Bemporad65,Sandorfi83}. The polarisation properties of high energy photons produced by the HICS process are of fundamental importance in nuclear physics applications, for instance in the study of abnormal parity components in the deuteron wavefunction \cite{GriRek83,RavRam85}. The polarisation state of the emitted HICS photon, as a function of initial polarisation states, for a circularly polarised external field, was studied by \cite{GriRek83,Tsai93}.\\

In experimental work on the HICS process, a 1.053 $\mu$m laser focused to a 5 $\mu$m spot size to generate a peak laser intensity of $\sim 10^{18}\,\text{Wcm}^{-2}$ interacted with electrons resulting in a longitudinal shift in scattered electron momenta due to multi-photon contributions from the laser field. Also observed was the predicted electron mass shift due to the presence of the external field \cite{Meyerhofer95,Meyerhofer96}. The HICS process and another strong field process, the one photon pair production process, were studied at SLAC in the 1990s by interacting a Nd:glass laser of peak intensity $0.5\times10^{18} \text{ W cm}^{-2}$ with a 46 GeV electron beam \cite{Bamber99}\\

The prospect in the near future of extreme light sources producing laser pulses with intensities exceeding those of previous experiments has prompted new theoretical work in pulsed background fields. The HICS process was studied in a circularly polarised field with pulse envelope given by a general function \cite{NarFof96}. The trident process was studied using the Weizsacker Williams approximation \cite{BulMcd00} and a complete calculation using pulsed laser field was performed by \cite{Ildert11}. One photon pair annihilation in a pulsed laser field with specific attention to operating parameters at the future XFEL was also studied recently \cite{IldJoh11}. \\

The HICS process is also fundamentally important in accelerator physics where it is known as the Beamstrahlung and where it constitutes the main source of background photons due to beam-beam effects at the interaction point of a collider \cite{YokChe91}. The external field in the Beamstrahlung is provided by an oncoming relativistic charge bunch and is a constant crossed electromagnetic field described by a 4-potential of infinite period. Interactions within a constant crossed field permit the radiation of vanishingly soft photons and the differential transition rate is divergent \cite{Hartin09}. \\

In order to study the general properties of the HICS process it would be useful to have an analytic expression valid for all external fields for which exact solutions of the Dirac equation are possible and of which those external fields described hitherto in this section are specific examples. The advantage of such a general analytic expression is that not only known expressions for particular external fields could be written down immediately from a single source, but also the HICS transition probability for new external fields not yet studied theoretically would be readily available. The purpose of this paper, then, is to derive an expression for the HICS transition rate in a general external plane wave electromagnetic field and to illustrate its usage with particular examples.\\

%\vspace{0.8cm}
%\begin{minipage}{0.45\textwidth}
%\centering
%\includegraphics[width=0.7\textwidth]{../../myfigs/1storder_vertex.pdf}
%\captionof{figure}{\bf The HICS process Feynman diagram.}\label{fig:hicsfeyn}
%\label{coord}
%\end{minipage}\hspace{0.05\textwidth}
%\begin{minipage}{0.45\textwidth}
%\centering
%\includegraphics[width=0.85\textwidth]{../../myfigs/HICS_ang.pdf}\label{hicsang}
%\captionof{figure}{\bf The HICS 4-momenta and coordinate system.}
%\end{minipage}
%\vspace{0.8cm}

\begin{figure}[h!] 
%\centerline{\includegraphics[width=0.25\textwidth]{../../myfigs/1storder_vertex.pdf}}
\centerline{\includegraphics[width=0.25\textwidth]{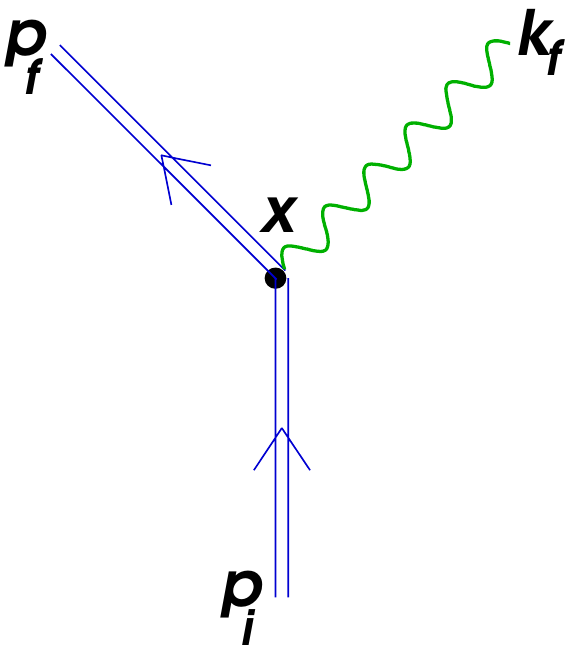}}
\caption{\bf The HICS process Feynman diagram.}\label{fig:hicsfeyn}
\end{figure} 

\section{The Volkov Solution for a general plane wave external field} 

\sloppy
In terms of notation, we employ a metric with signature $(1,-1,-1,-1)$ and write the time and 3-vector part of a 4-vector by $x^\mu=(x^0,\vec{x})$. Conventionally, we work in natural units $\hbar=c=1$. \\

\begin{figure}[h!] 
%\centerline{\includegraphics[width=0.30\textwidth]{../../myfigs/HICS_ang.pdf}}
\centerline{\includegraphics[width=0.30\textwidth]{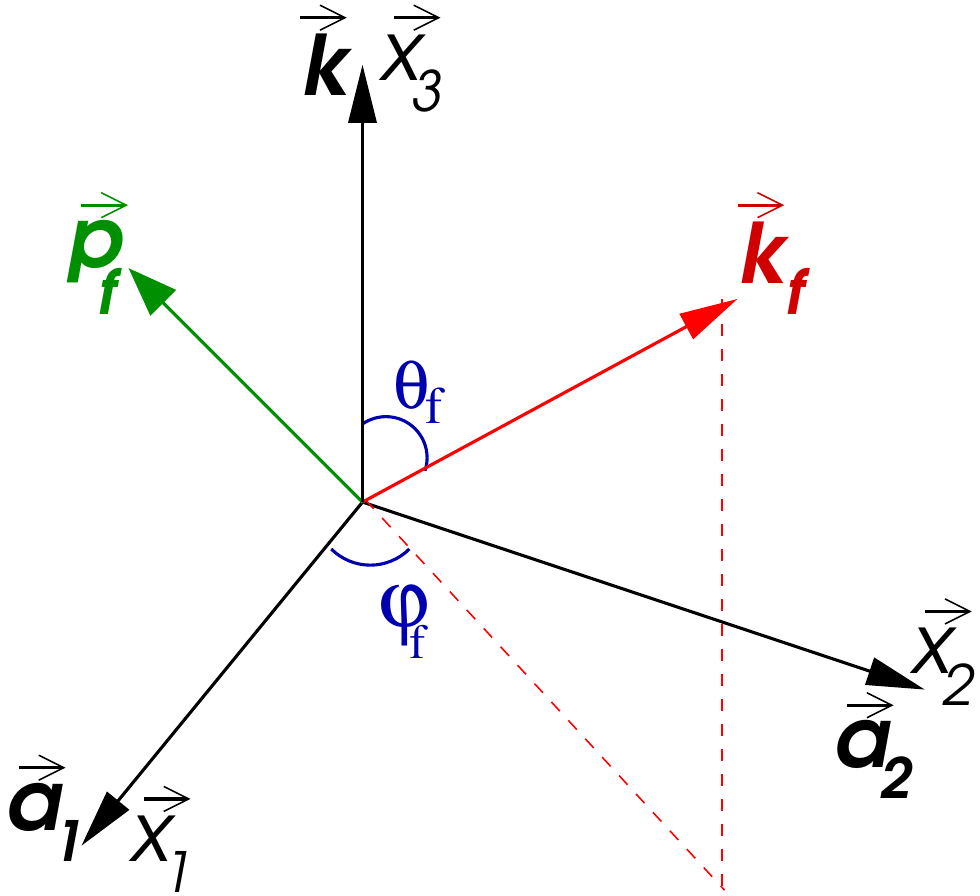}}
\caption{\bf The HICS coordinate system.}\label{fig:hicsang}
\end{figure}

A general plane wave electromagnetic field with wavevector $\vec{k}$ directed along the $\vec{x}_3$ axis of a Cartesian coordinate system (figure \ref{fig:hicsang}) is described by a 4-potential $A^e_\mu$ which can be decomposed into two 4-vectors $a_{1\mu}=(0,\vec{a}_1),\,a_{2\mu}=(0,\vec{a}_2)$ whose 3 vector parts are parallel to the $\vec{x}_1,\vec{x}_2$ axes and whose co-factors $A^{e,1}_i,A^{e,2}_i$  are general functions of arbitrary period described by a sum of $N$ terms, 

\begin{gather}\label{Eq:GeneralA}
A^e_\mu(k\cdot x) \equiv\sum\limits^N_{i=1} a_{1\mu} A_i^{e,1}(k\cdot x)+a_{2\mu} A_i^{e,2}(k\cdot x)
\end{gather}

The strength of the external field is an important factor determining the extent of multi-photon interactions, and is characterised by the dimensionless parameter $\nu$ which is a ratio (in natural units) of the fermion charge $e$, mass $m$ and field potential,

\begin{gather}\label{Eq:nudef}
\nu\equiv\frac{ea}{m} \quad,\quad a\equiv |\vec{a}_1|=|\vec{a}_2|
\end{gather}

In the Furry picture the Dirac equation containing the 4-potential $A^e_\mu$ for a fermion of mass $m$ and momentum $p=(\epsilon_p,\vec{p})$,

\begin{align}\label{eq:QEDbdiraceqn}
(i\slashed{\partial}+A^e_{\mu}-m) \Psi_p^V=0
\end{align}

can be solved exactly, resulting in wavefunctions $\Psi_p^V$ first obtained by Volkov \cite{Volkov35}. These Volkov solutions are a product of the normal free fermion solution including the bispinor $u_p$ and normalisation factor, with an extra phase $S_p$ and a term containing $\slashed{A^e}\slashed{k}$,

\begin{gather}\label{eq:Volkov}
 \Psi_p^V(k\cdot x)= \,\mfrac{1}{\sqrt{(2\pi)^3 2\epsilon_p}}\,E_p(k\cdot x)\, u_p \\
\text{where}\hspace{0.5cm} E_p(k\cdot x)\equiv\lp 1-\mfrac{e\slashed{A}^e\slashed{k}}{2(k\cdot p)}\rp \hspace{3cm} \notag\\
\hspace{2cm}\centerdot\exp{\lp-i\lp p+\mfrac{e^2a^2\xi}{2(k\cdot p)}k\rp\cdot x-i S_p(k\cdot x)\rp}\notag \\ 
S_p(k\cdot x)\equiv\int_{0}^{(k\cdot x)}\ls \mfrac{e(A^{e}(\phi)\cdot p)}{(k\cdot p)}-\mfrac{e^2A^e(\phi)^2}{2(k\cdot p)}-\mfrac{e^2a^2\xi}{2(k\cdot p)}\rs d\phi \notag
\end{gather}

In the Volkov solution above we have anticipated a quasi-momentum term containing the parameter $\xi$ which has a linear dependence on the scalar product $(k\cdot x)$. The quasi-momentum is physically due to the average change in motion of the electron embedded in an external, oscillatory external electromagnetic field. The precise form of $\xi$ depends on the specific external field the HICS process takes place in. \\

%Volkov-type solutions have also been found for the case of two external field potentials $A^e=A^e_{a\mu}+A^e_{b\mu}$ for the orthogonal case in which $(A^e_a\cdot A^e_b)=0$ The solution of the Dirac equation in this case is a product of Volkov solutions \cite{Lyulka75,Pady06}. \\

The Volkov wavefunctions are used within the normal S-matrix theory to study interactions between the fermion embedded in the external field and the quantised boson field.\\ 

In order to calculate a transition probability, the dependence on the scalar product $(k\cdot x)$ must be simplified. The natural way of doing this is to write down a matrix element from a Feynman diagram (figure \ref{fig:hicsfeyn}, with double straight lines representing Volkov wavefunctions) and gathering the dependence on space-time around the diagram vertices. The result is a modified vertex $\gamma^e_\mu$ which can be transformed from position space to momentum space using a Fourier transform for a function of arbitrary period $2\pi L$, 

\begin{gather}\label{Eq:Volkovgen}
\gamma_{\mu}^e(p_f,p_i,k\cdot x) \equiv \bar E_{p_f}(k\cdot x)\gamma_\mu E_{p_i}(k\cdot x) \notag\\[6pt]
 \gamma_{\mu}^e(p_f,p_i,k\cdot x) = \sum\limits_{r=-\infty}^{\infty}\int_{-\pi L}^{\pi L} \frac{d\phi_v}{2\pi L} \hspace{2cm}\\[4pt]
 \hspace{2cm}\centerdot \exp{\lp i\mfrac{r}{L}\ls \phi_v-(k\cdot x)\rs\rp} \gamma_{\mu}^e(p_f,p_i,\phi_v) \notag 
\end{gather}

\section{The HICS transition probability for a general external field} 

The HICS transition probability is calculated in the usual fashion, writing down a matrix element, squaring it, performing a trace and integrating over the phase space.\\

The matrix element for the HICS process can be written down directly from the Feynman diagram in figure \ref{fig:hicsfeyn}. The matrix element is essentially the Fourier transform of the Volkov vertex with a Dirac delta function expressing the conservation of momenta, including a contribution $\frac{r}{L}k$ from the external field and the quasi-momentum term. Denoting initial and final fermion 4-momenta $(\varepsilon_i,\vec{p}_i)$,\,$(\varepsilon_f,\vec{p}_f)$ and the radiated photon 4-momentum $(\omega_f,\vec{k}_f)$, the matrix element is

\begin{multline}\label{eq:HICSmatel}
M_{fi}= \mfrac{-ie}{\sqrt{16\pi\epsilon_i\epsilon_f\omega_f}}
\sum\limits_{r=-\infty}^{\infty}\int_{-\pi L}^{\pi L} \mfrac{d\phi_v}{2\pi L} \\ 
\centerdot \bar u_{p_f} \,\gamma^{e\mu}(p_f,p_i,\phi_v)\varepsilon_\mu(k_f) \,u_{p_i} \exp{\lp i\frac{r}{L}\phi_v\rp}\\[4pt]
\centerdot
\delta\lp p_f+k_f-p_i-\lp\mfrac{r}{L}-\mfrac{e^2a^2\xi(k\cdot k_f)}{2(k\cdot p_i)(k\cdot p_f)}\rp k\rp
\end{multline}

Squaring the matrix element results in a double summation (over indicies $r,r'$). The subsequent product of delta functions will yield zero for $r\neq r'$ and otherwise the product of space-time volume $VT$ in which the process takes place and a single delta function in index $r$. The spin sums yield the usual projection operators $\Lambda_p=\slashed{p}+m$ and trace, and there is a double integration over variables $\phi_v,\phi_w$

\begin{multline}\label{eq:HICSsqmat}
\sum\limits_{if}^{}\frac{|M_{fi}|^2}{VT}=-\mfrac{e^2}{32\pi\epsilon_i\epsilon_f\omega_f}
\sum\limits_{r=-\infty}^{\infty}\int_{-\pi L}^{\pi L} \mfrac{d\phi_v\,d\phi_w}{(2\pi L)^2} \\
\centerdot\Tr{\left[\Lambda_{p_f}\gamma^{e\mu}(p_f,p_i,\phi_v)\Lambda_{p_i}\gamma^e_\mu(p_i,p_f,\phi_w)\right]} \\[4pt]
\centerdot\exp{\lp i\mfrac{r}{L}(\phi_v-\phi_w)\rp}\\
\centerdot \delta\lp p_f+k_f-p_i-\lp\mfrac{r}{L}-\mfrac{e^2a^2\xi(k\cdot k_f)}{2(k\cdot p_i)(k\cdot p_f)}\rp k \rp
\end{multline}

The trace calculation is lengthy but straightforward. A simplification was obtained by grouping terms in order to separate the derivative with respect to the integration variables $\phi_v,\phi_w$. The result of the trace is,

\begin{gather}\label{eq:HICStrace}
4 m^2 \lb 2 -\mfrac{e^2}{2m^2} \lp \mfrac{(k\cdot p_i)}{(k\cdot p_f)} + \mfrac{(k\cdot p_f)}{(k\cdot p_i)} \rp
\ls A^e(\phi_v)-A^{e}(\phi_w)\rs^2 \rb \notag\\[4pt]
\centerdot \exp\left(i\left[S_{p_f}(\phi_v)-S_{p_i}(\phi_v)\right]-i\ls S_{p_f}(\phi_w)-S_{p_i}(\phi_w)\rs\rp
\end{gather}

The integration over phase space is carried out as usual with the aid of the delta function. It is convenient to transform to integrations over 4-vectors and perform the integration over the final electron momentum leaving the integration over the photon 4-momentum,

\medskip
\begin{multline}\label{eq:PIident}
\int\mfrac{d\vec{k_f}\,d\vec{p_f}}{4\epsilon_f\,\omega_f} \;\delta\lp p_f+k_f-p_i-\lp\mfrac{r}{L}-\mfrac{e^2a^2\xi(k\cdot k_f)}{2(k\cdot p_i)(k\cdot p_f)}\rp k\rp \\
=\int d^4k_f\; \delta(k_f^2) \\
\centerdot\delta\lp 2\lp\mfrac{r}{L}-\mfrac{e^2a^2\xi(k\cdot k_f)}{2(k\cdot p_i)(k\cdot p_f)}\rp (k\cdot p_f)-2(p_i\cdot k_f) \rp
\end{multline}\medskip

In performing the integration over final photon momentum it proved useful to transform to light-cone coordinates,

\begin{gather}\label{eq:lightcone}
d^4k_f \rightarrow \mfrac{1}{2} dk_f^{+}\,dk_f^{-}\,dk_f^1\,dk_f^2 \\
\quad\text{where}\quad k_f^{+}=k_f^{0}+k_f^{3} \; , \; k_f^{-}=k_f^{0}-k_f^{3} \notag
\end{gather}

Choosing the reference frame in which $\vec{p}_i=0$ for simplicity, the integration over $dk_f^{+}$ is performed using the second delta function in equation \eqref{eq:PIident}. The integration over $dk_f^{-}$ is transformed to one over $u=(k\cdot k_f)/(k\cdot p_f)$ and is retained. The interim result for the transition probability is

\begin{multline}\label{eq:HICSWinterim}
W=-\mfrac{e^2m}{4\pi} \sum\limits_{r=-\infty}^{\infty}\int_{-\pi L}^{\pi L} \mfrac{d\phi_{v}\,d\phi_{w}}{(2\pi L)^2} \int_0^{\infty} \nths\mfrac{du}{(1+u)^2} \int_{-\infty}^{\infty} dk_f^1 \; dk_f^2 \\
\centerdot \delta\lp 2\lp\mfrac{r}{L}-\mfrac{m^2\nu^2u}{2(k\cdot p_i)}\xi\rp\mfrac{(k\cdot k_f)(k\cdot p_f)}{(k\cdot p_i)}-k_f^{-\,2}-k_f^{1\,2}-k_f^{2\,2}\rp \\[8pt]
\centerdot\lb 2-\mfrac{e^2}{2m^2}  \mfrac{1+(1+u)^2}{1+u} 
\ls A^e(\phi_v)-A^{e}(\phi_w)\rs^2\rb \\[8pt]
\centerdot\exp\lp i\ls\lp\mfrac{r}{L}-\mfrac{m^2\nu^2 u}{2(k\cdot p_i)}\xi\rp (\phi_{v}-\phi_w) \rd\rd \\
 \ld\ld -\mfrac{e}{(k\cdot p_f)}\int_{\phi_w}^{\phi_v}\nths (k_f\cdot A^{e}(\phi))d\phi
-\mfrac{e^2u}{2(k\cdot p_i)}\int_{\phi_w}^{\phi_v}\nths A^e(\phi)^2\,d\phi\rs\rp
\end{multline}

In simplifying the interim result above for the HICS transition probability (equation \eqref{eq:HICSWinterim}) there are two ways of proceeding. The first way, which proves useful for constant external fields, is to transform the delta function to an exponential and then to use the Fourier transform of a $2\pi L$ period function $f(\phi)$ to remove the dependence on $r$,

\begin{gather}\label{Eq:FTarbitrary}
\sum\limits_{r=-\infty}^{\infty}\int_{-\pi L}^{\pi L} \frac{d\phi}{2\pi L}
\exp{\lp i\mfrac{r}{L}\ls \phi-\lambda\rs\rp} f(\phi) = f(\lambda) 
\end{gather}

The integrations over $dk_f^1\,dk_f^2$ appear as Gaussian integrals and are readily carried out to obtain the HICS transition probability which we label $W_{\text{HICS}}^{\text{1st}}$. Defining auxillary functions,

\begin{gather}
F_1(\phi_v,\phi_w)\equiv -\mfrac{e^2}{m^2}\ls A^e(\phi_v)-A^{e}(\phi_w) \rs^2 \notag\\[4pt]
F_2(\phi_v,\phi_w)\equiv \mfrac{e}{m}\lv \int_{\phi_w}^{\phi_v}\nths A^{e}(\phi)d\phi \rv \\
F_3(\phi_v,\phi_w)\equiv -\mfrac{e^2}{m^2}\int_{\phi_w}^{\phi_v}\nths A^e(\phi)^2\,d\phi \notag
\end{gather}
 
we have,

\begin{multline}\label{eq:HICSWfinal1st}
W_{\text{HICS}}^{\text{1st}}=-\mfrac{e^2m}{4\pi}\int_{-\pi L}^{\pi L} \mfrac{d\phi_w}{2\pi L} \int_0^{\infty} \nths\mfrac{du}{(1+u)^2} \int_{-\infty}^{\infty} \mfrac{i\pi d\lambda}{2\lambda}\\
\centerdot \lb 2+\mfrac{1+(1+u)^2}{2(1+u)} F_1(\lambda\!+\!\phi_w,\phi_w)\rb \\
\centerdot\exp\lp i\mfrac{m^2u}{2(k\cdot p_i)}\lp\lambda-\mfrac{F_2(\lambda+\phi_w,\phi_w)^2}{\lambda}+F_3(\lambda\!+\!\phi_w,\phi_w)\rp \rp
\end{multline}

The second route from the HICS interim transition probability, which proves useful for oscillatory external fields, is to carry out the $dk_f^1\,dk_f^2$ integrations in polar coordinates, $zdz\,d\theta$. The delta function from equation \eqref{eq:HICSWinterim} becomes,

\begin{gather}
\delta\lp z^2-2\lp\mfrac{r}{L}-\mfrac{m^2(1+\nu^2\xi)u}{2(k\cdot p_i)}\rp\mfrac{(k\cdot k_f)(k\cdot p_f)}{(k\cdot p_i)} \rp
\end{gather}

and can be used to perform the integration over $dz$, thereby setting a range for the summation over $r$ and restricting the upper limit of the $du$ integration. The integration over $d\theta$ results in a Bessel function. It proves useful to make a change of integration variables from $d\phi_v\,d\phi_w$ to $d\phi_{-}\,d\phi_{+}$. The end result is labelled $W_{\text{HICS}}^{\text{2nd}}$ and is

\begin{multline}\label{eq:HICSWfinal2nd}
W_{\text{HICS}}^{\text{2nd}}=-\mfrac{e^2m}{2\pi} \sum\limits_{r=-\infty}^{\infty}\Theta(r)\int_{-\pi L}^{\pi L} \mfrac{d\phi_{-}\,d\phi_{+}}{(2\pi L)^2} \int_0^{u_r} \nths\mfrac{du}{(1+u)^2} \\
\centerdot \lb 2+\mfrac{1+(1+u)^2}{2(1+u)} F_1(\phi_v,\phi_w)\rb \text{J}_0\biggl( 2zF_2(\phi_v,\phi_w)\biggr) \\
\centerdot\exp\lp i\lp\mfrac{r}{L}-\mfrac{m^2\nu^2 u}{2(k\cdot p_i)}\xi\rp 2\phi_{-}+i\mfrac{m^2u}{2(k\cdot p_i)}F_3(\phi_v,\phi_w) \rp
\end{multline}

where,

\begin{gather*}
z\equiv \ls\mfrac{m^2u}{2(k\cdot p_i)}\lp\mfrac{r}{L}-\mfrac{m^2(1+\nu^2\xi)u}{2(k\cdot p_i)}\rp\rs^{1/2} \\[4pt]
u_r\equiv 2\mfrac{(k.p_i)}{m^2(1+\nu^2\xi)}\mfrac{r}{L}\\
\phi_{-}=\frac{1}{2}(\phi_v-\phi_w), \quad \phi_{+}=\frac{1}{2}(\phi_v+\phi_w) \\
\Theta(r)=0 \;\text{for}\; r\leq 0, \quad \Theta(r)=1 \;\text{for}\; r>0
\end{gather*}

We now have analytic representations for the HICS transition probability for a plane-wave electromagnetic field of general form. It should be now possible to derive existing results from the literature and to write down new ones for external field forms not yet considered.

%---------------------------------------------------------------------
% for dk_f_x and dk_f_y in cartesian coordinates
%---------------------------------------------------------------------
%A considerable simplification is obtained by transforming the delta function to an exponential and performing the integrations over $k_f^x,k_f^y$ after extracting their dependence from the $S_p$ functions. The final result is
%
%\medskip
%\begin{gather}\label{eq:HICSWfinal}
%W=\mfrac{e^2m}{4\pi}\int_{-\pi L}^{\pi L} \mfrac{dw}{2\pi L} \int_0^{\infty} \nths\mfrac{du}{(1+u)^2} \int_{-\infty}^{\infty} \mfrac{i\;d\lambda}{2\lambda}\left\{ 2+\mfrac{1+(1+u)^2}{1+u} F_1(\lambda,w)\right\} \,\exp\left(i\;\mfrac{u\,m^2}{2(k\cdot p_i)}F_2(\lambda,w)\right) \notag\\[10pt]
% F_1(\lambda,w)\equiv \mfrac{e^2}{2m^2}\ls A^e(\lambda +w)-A^{e\ast}(w) \rs^2 \\[8pt]
% F_2(\lambda,w)\equiv \lambda-\mfrac{e^2}{m^2} \left[ \int_w^{\lambda+w}\nths A^e(\phi)^2\,d\phi 
%+ \mfrac{1}{\lambda} \lp \int_w^{\lambda+w}\nths A^{e}(\phi)d\phi \rp^2\rs \notag 
%\end{gather}\medskip
%-----------------------------------------------------------------------

\section{The HICS transition probability for specific external fields} 

To illustrate the general expression for the HICS transition probability obtained in the last section, two analytic results and a numerical result are obtained for particular external fields of practical interest. The specific analytic form for the HICS transition probability can be obtained from either equation \eqref{eq:HICSWfinal1st} or \eqref{eq:HICSWfinal2nd} after substitution of the particular 4-potential $A_e$ of period $2\pi\omega L$ into the auxiliary functions $F_1,F_2,F_3$, choosing a value for the quasi-momentum parameter $\xi$ in order to simplify the analytic expression, and by performing further integrations where possible.

\subsection{A circularly polarised external field}

Experiments investigating strong field effects are usually performed with intense lasers which commonly provide a circularly polarised electromagnetic field. Such a field is described by the 4-potential and resultant auxiliary functions,

\begin{gather}\label{eq:Acircpol}
A^e_\mu(\phi)=a_{1\mu} \cos(\phi)+ a_{2\mu} \sin(\phi),\quad L=1,\quad\xi=1 \notag\\
F_1(\phi_v,\phi_w) = 2\nu^2(1-\cos2\phi_{-}) \notag\\
F_2(\phi_v,\phi_w) =  2\nu\lv \sin\phi_{-} \rv \\ 
F_3(\phi_v,\phi_w) = 2\nu^2\phi_{-}\notag
\end{gather}

The quasi-momentum parameter $\xi=1$, results in a simplified exponential dependence in the second general form of the HICS transition rate, equation \eqref{eq:HICSWfinal2nd}. There is no dependence in the transition probability integrand on the variable $\phi_{+}$ rendering its integration trivial. The $d\phi_{-}$ integration can be carried out with the aid of the Bessel function identity (\cite{Watson22} \S 2.6), 

\begin{gather}\label{eq:Bessrel}
\J_r(z)^2=\mfrac{1}{2\pi} \int^{\pi}_{-\pi} d\phi_{-}\;\J_0(2z\sin\phi_{-})\,\exp\lp i\,2r\phi_{-}\rp
\end{gather}

The final result for the HICS transition probability for a circularly polarised external field is the same as that found in the literature (\cite{BerLifPit82}, equation 101.15),

\begin{multline}\label{eq:Wcirc}
W_{\text{HICS}}^{\text{circ}}=-\mfrac{e^2m}{2\pi}\sum\limits_{r=1}^{\infty}\int_0^{u_r} \nths\mfrac{du}{(1+u)^2} \\
\centerdot\ls 2\J_{r}^2(z)-\mfrac{\nu^2}{2}\mfrac{1+(1+u)^2}{1+u} 
\lp\J_{r+1}^2(z)+\J_{r-1}^2(z)-2 \J_{r}^2(z)\rp\rs \\
z\equiv 2\nu\ls\mfrac{m^2u}{2(k\cdot p_i)}\lp r-\mfrac{m^2(1+\nu^2)u}{2(k\cdot p_i)}\rp\rs^{1/2},u_r\equiv \mfrac{2(k.p_i)\,r}{m^2(1+\nu^2)}
\end{multline}

The above expression for the HICS transition probability has been extensively numerically studied in the literature. However it is worth noting that a numerical study could proceed directly from the general expression for the HICS transition probability in equation \eqref{eq:HICSWfinal2nd}.
 
\subsection{A constant crossed external field}

The Beamstrahlung results from charges within a particle beam at the interaction point of a collider decelerating and radiating within the strong field of an oncoming, ultrarelativistic charge bunch. Such an external field can be described by an infinite period, constant crossed electromagnetic field. The first form of the general HICS transition probability (equation \eqref{eq:HICSWfinal1st}) will allow us to directly write the known result for the HICS transition probability in this field. The 4-potential and resultant auxiliary functions are, 

\begin{align}\label{eq:Abeamstr}
A^e_\mu(\phi)=a_{1\mu} \phi, \quad L=\infty, \quad \xi=0 \notag\\
F_1(\lambda\!+\!\phi_w,\phi_w) = \nu^2\lambda^2 \notag\\[4pt]
F_2(\lambda\!+\!\phi_w,\phi_w) =  \mfrac{\nu}{2}\lambda^2 \\ 
F_3(\lambda\!+\!\phi_w,\phi_w) = \mfrac{\nu^2}{3}\,\lambda^3 \notag
\end{align}

There is no dependence on the quasi-momentum parameter $\xi$, reflecting the fact that the electron does not oscillate in a constant field. Substitution of the infinite period $L=\infty$, poses no problem since there is no dependence of the integrand on the $d\phi_w$ integration and,

\begin{gather}
\int_{-\pi L}^{\pi L} \mfrac{d\phi_w}{(2\pi L)}=1
\end{gather}

%----------A useful idenitity-----------------------
%Starting from
%
%\begin{gather*}\label{eq:identprep}
%\int_{-\infty}^{\infty} dk_f^1\,dk_f^2\, \delta \lp k_f^{1\,2}+k_f^{2\,2}-a \rp \e^{i x k_f^1+i y k_f^2}
%\end{gather*}
%
%and performing the integrations $dk_f^1\,dk_f^2$ in either Cartesian or polar coordinates, the identity
%
%\begin{gather}\label{eq:newident}
%\Theta(a)\J_0 \lp \sqrt{ab} \rp= \mfrac{i\pi}{2} \int_{-\infty}^{\infty} \mfrac{d\lambda}{\lambda}  \e^{-ia\lambda -i b/4\lambda} ,\quad b>0
%\end{gather}
%
%is obtained.\\

Substitution of the auxillary functions of equation \eqref{eq:Abeamstr} yields an interim HICS transition probability for a constant crossed external field,

\begin{multline}\label{eq:Wconstinterim}
W_{\text{HICS}}^{\text{const}}=-\mfrac{e^2m}{4\pi}\int_0^{\infty} \nths\mfrac{du}{(1+u)^2} \int_{-\infty}^{\infty} i\;d\lambda \\
\centerdot\ls \mfrac{1}{\lambda}-\mfrac{\nu^2}{4}\mfrac{1+(1+u)^2}{1+u} \lambda\rs \\
\centerdot\exp\lp i\;\mfrac{u\,m^2}{2(k\cdot p_i)}\lp\lambda+\mfrac{1}{12}\nu^2\lambda^3\rp\rp 
\end{multline}

We need now only to rescale $\lambda$ to get the integral representation with respect to $\lambda$ of an Airy function. The final form of the HICS tranition probability for a constant crossed field, which is the same as that appearing in the literature \cite{Ritus72}, is

\begin{multline}\label{eq:Wconstfinal}
W_{\text{HICS}}^{\text{const}}=-\mfrac{e^2m}{2}\int_0^{\infty} \nths\mfrac{du}{(1+u)^2} \ls \int dz +\mfrac{1+(1+u)^2}{z\,(1+u)} \frac{d}{dz}\rs \Ai(z) \\[4pt]
\text{where} \quad\quad z\equiv \lp\mfrac{u\,m^2}{\nu(k\cdot p_i)}\rp^{2/3} \hspace{2cm}
\end{multline}

\subsection{Two phase offset circularly polarised external fields}
In at least two practical endeavours, namely laser-plasma acceleration of charged particle bunches \cite{ShvFis02} and laser confinement fusion experiments \cite{NakMim04}, multiple intense laser interactions with matter take place. In that spirit we examine the HICS transition probability for the simple multi-field case of two co-linear circularly polarised external fields of equal energy and intensity, separated by a variable phase difference denoted by $\chi$. The 4-potential of the combination of field components and the resultant auxiliary functions are,

\begin{gather}\label{eq:Anew1}
A^e_\mu(\phi)=a_{1\mu} (\cos\phi+\cos(\phi\!+\!\chi))+a_{2\mu} (\sin\phi+\sin(\phi\!+\!\chi)) \notag\\
L=1,\quad \xi= 2(1+\cos\chi) \notag\\
F_1(\phi_v,\phi_w) = 8\nu^2\cos^2\nthn\lp\mfrac{\chi}{2}\rp(1-\cos2\phi_{-}) \\
F_2(\phi_v,\phi_w) = 4\nu\lv\cos\lp\mfrac{\chi}{2}\rp \sin\phi_{-} \rv \notag\\
F_3(\phi_v,\phi_w) = 4\nu^2(1+\cos\chi)\phi_{-} \notag
\end{gather}

The quasi-momentum parameter $\xi$ in this case depends on the phase difference between the two fields. After substitution of the auxillary functions into the second general expression for the HICS transition probability a transition probability for two phase-offset circularly polarised fields, $W_{\text{HICS}}^{\text{po}}$ is obtained. Apart from the appearance of the phase difference $\chi$ the analytic form of the transition probability resembles that for a single circularly polarised field,

\begin{multline}\label{eq:Wphase}
W_{\text{HICS}}^{\text{po}}(\chi)=-\mfrac{e^2m}{2\pi}\sum\limits_{r=1}^{\infty}\int_0^{u_r} \nths\mfrac{du}{(1+u)^2} \\
\centerdot \biggl[ 2\J_{r}^2(z)-2\nu^2\cos^2\nthn\lp\mfrac{\chi}{2}\rp\mfrac{1+(1+u)^2}{1+u} \\
\centerdot\lp\J_{r+1}^2(z)+\J_{r-1}^2(z)-2 \J_{r}^2(z)\rp \biggr]
\end{multline}

where

\begin{gather*}
z\equiv 4\nu\cos\mfrac{\chi}{2}\ls\mfrac{m^2u}{2(k\cdot p_i)}\lp r-\mfrac{m^2(1+2\nu^2(1+\cos\chi))u}{2(k\cdot p_i)}\rp\rs^{1/2} \\
u_r\equiv \mfrac{2(k.p_i)r}{m^2(1+2\nu^2(1+\cos\chi))} \hspace{2cm}
\end{gather*}

\begin{figure}[h!] 
%\centerline{\includegraphics[width=0.45\textwidth]{../../myplots/2circpol_samehel_Wvphase.pdf}}
\centerline{\includegraphics[width=0.45\textwidth]{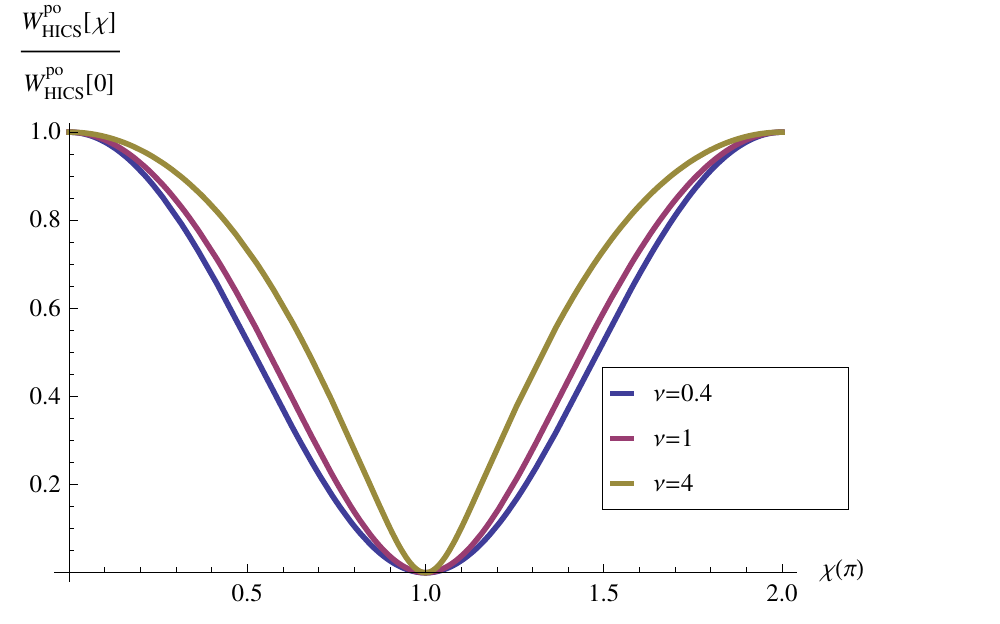}}
\caption{\bf HICS transition probability for 2 circularly polarised fields of momentum 5.11 MeV and phase difference $\chi$}\label{fig:hicscirc1}
\end{figure}

A numerical evaluation of equation \eqref{eq:Wphase} is straightforward since the summand converges quickly and the oscillatory nature of the integrand is inhibited. Figure \ref{fig:hicscirc1} shows the variation of the HICS transition probability with phase offset for different external field intensities. The HICS transition probability is maximal when the two fields are in phase, and vanishes when the phase difference is $\pi$ and the two fields destructively interfere 

\section{Conclusion} 
In this paper we have set out to write down a new analytic expression for the transition probability of the High Intensity Compton Scattering in an external field of general form. Two equivalent expressions for the HICS transition probability were obtained. One of these general expressions is suitable for obtaining simple, specific analytic expressions for constant external fields of infinite period and a vanishing quasi-momentum term, The other general expression is suitable for oscillatory external fields. \\

Exact Volkov solutions of the electron in a general plane wave electromagnetic field whose 4-potential is a sum of N co-linear components have been used. By choosing such a representation for a general plane wave, any waveform that can be expressed by a Fourier series can be studied using general forms of the HICS transition probability written down in equations \eqref{eq:HICSWfinal1st} and \eqref{eq:HICSWfinal2nd}. \\

The validity of the general HICS transition probability expressions was established by substituting in particular 4-potentials - that of a circularly polarised and of a constant crossed electromagnetic field - to obtain the same analytic forms of the HICS transition probability found in the literature. Specific numerical studies can, however, be launched directly from the general HICS transition probability expressions which is particularly useful for external fields for which no simple analytic expression for the transition probability is possible. \\

A multi-external field case, in which the HICS process takes place within two co-linear circularly polarised fields of equal intensity and energy separated by a phase difference, was examined analytically and numerically. As the phase difference reaches $\pi$, the two fields destructively interfere and the HICS transition probability vanishes. \\

New high intensity laser facilities are being planned and built which will provide the means to experimentally test physics processes in strong electromagnetic fields. A variety of experimental conditions involving different configurations and combinations of external fields will be possible. In the case of experimental tests of the HICS process we have provided a general expression for the HICS transition probability from which, no matter what the experimental configuration, specific analytic expressions and numerical values can be quickly generated. 

\bibliographystyle{unsrt}
\bibliography{../hartin_bibliography}

%\begin{thebibliography}{9} 
%\bibitem{HartinPhD} 
% Hartin A 2006 {\it PhD thesis} University of London 
%\bibitem{Bamb99} 
%  Bamber C et al 1999 {\it Phys Rev D} {\bf 60}(9) 092004 
%\bibitem{Volk35} 
%  Volkov D M 1935 {\it Z Phys} {\bf 94} 250 
%\bibitem{Zeld67} 
%  Zeldovich Y B 1967 {\it Sov Phys JETP} {\bf 24} 1006 
%\bibitem{Ritus72} 
%  Ritus V I 1972 {\it Ann.\ Phys.\ D} {\bf 69} 555-582 
%\bibitem{BecMit76} 
%  Becker W, Mitter H 1976 {\it J Phys A} {\bf 9}(12) 2171 
%\bibitem{Ritus70} 
%  Ritus V I 1970 {\it Sov. Phys. JETP} {\bf 30}(6) 1181 
%\end{thebibliography} 
 
\end{document}